\documentclass[doublecol]{epl2}

\title{Possible Experience: from Boole to Bell}
\shorttitle{Possible Experience: from Boole to Bell} %Insert here a short version of the title if it exceeds 70 characters

\institute{
  \inst{1} Beckman Institute, Department of Electrical and Computer Engineering and Department of Physics\\
           University of Illinois, Urbana, Il 61801\\
  \inst{2} Institute for Advanced Simulation, J\"ulich Supercomputing Centre,
           Research Centre J\"ulich\\ D-52425 J\"ulich, Germany, EU\\
  \inst{3} Department of Applied Physics, Zernike Institute for Advanced Materials,
           University of Groningen\\ Nijenborgh 4, NL-9747 AG Groningen, The Netherlands, EU
}

\author{K. Hess\inst{1}\thanks{E-mail: k-hess@illinois.edu}
 \and K. Michielsen\inst{2}\thanks{E-mail: k.michielsen@fz-juelich.de} 
 \and H. De Raedt\inst{3}\thanks{E-mail: h.a.de.raedt@rug.nl}
}
\shortauthor{K. Hess, K. Michielsen and H. De Raedt}

\received{1 July 2009}
\acceptedinfinalform{11 September 2009}
\onlinepub{9 October 2009}


\def\revision#1{#1}

\abstract{%
Mainstream interpretations of quantum theory maintain that
violations of the Bell inequalities deny at least either realism or
Einstein locality. Here we investigate the premises of the Bell-type
inequalities by returning to earlier inequalities presented by Boole
and the findings of Vorob'ev as related to these inequalities. These
findings together with a space-time generalization of Boole's
elements of logic lead us to a completely transparent Einstein local
counterexample from everyday life that violates certain variations
of the Bell inequalities. We show that the counterexample suggests
an interpretation of the Born rule as a pre-measure of probability
that can be transformed into a Kolmogorov probability measure by
certain Einstein local space-time characterizations of the involved
random variables.
}%

\begin{document}

\maketitle

\section{Introduction}

We discuss models of Einstein-Podolsky-Rosen-Bohm
type~\cite{EPR35,BOHM57} of experiments as used by John
Bell~\cite{BELL64} when presenting his celebrated inequalities.
These experiments result in outcomes of two spin-values $\pm 1$ (in
units of $\hbar/2$) that in turn depend on certain magnet settings
${\bf a}, {\bf b}, {\bf c}...$ and have been linked to two-valued
functions $A_{\bf a}(\cdot), A_{\bf b}(\cdot), A_{\bf c}(\cdot) =
\pm 1$ by Bell and followers. Here $(\cdot)$ stands for the
dependency on some element of a set of mathematical representations
of elements of reality that do not depend on the magnet settings
${\bf a}, {\bf b}, {\bf c}...$. This latter fact of independence from
magnet settings was deduced by Bell from considerations of Einstein
locality and the (physically unjustified) assumption that the
elements of reality emanate exclusively from a distant source and
not from the measurement equipment (including the magnets). There
are numerous inequalities, delineated in the physics literature that
are related to Bell's functions $A_{\bf a}(\cdot), \ldots$. These
inequalities were first derived by Boole~\cite{BO1862} in a much
more general context. Here we discuss mainly a variation of the
inequalities as published by Leggett and Garg~\cite{LEGG85}, for
which we also have developed a transparent counterexample. More
complex counterexamples have been developed in the past for the more
elaborate inequalities~\cite{HESS01} but have remained largely
unappreciated because of their lack of transparency.
\revision{Nevertheless, strong movements critical of Bell's approach
continue to grow as demonstrated by contributions of Accardi, Fine,
Rastal, Khrennikov, Nieuwenhuizen and many others
~\cite{ACCA05,RAST81,FINE82a,THEO07,KHRE08}.
Here, however, we do not refer to non-Kolmogorovian
approaches~\cite{KHRE08} and we like to stress
that we also do not invoke detector inefficiencies or anything
related to fair sampling \cite{ADEN07}. Instead, our counterexample
is based on a more complete %space-time 
characterization of Boole's
logical elements. % and a corresponding change of Boole's inequalities.
}

The Leggett-Garg inequality reads:
\begin{equation}
A_{\bf a}(\cdot)A_{\bf b}(\cdot) + A_{\bf a}(\cdot)A_{\bf c}(\cdot)
+ A_{\bf b}(\cdot)A_{\bf c}(\cdot) \geq -1
.
\label{hla27n1}
\end{equation}
Inserting all possible values of $\pm 1$ for the functions $A(.)$
shows the correctness of this inequality. Because measurement
outcomes of Einstein-Podolsky-Rosen (EPR) experiments~\cite{ASPE82a}
(that are closely related to such two-valued functions $A(.)$) do
violate this inequality, it is commonly concluded that either
$(\cdot)$ can not stand for any element of reality and one must
therefore abandon realism or if it stands for an element of reality
it must depend on the magnet settings and thus violate Einstein
locality. There are, however, two important questions that have
never been answered satisfactorily. If $(\cdot)$ stands for an
element of reality, why does it have to appear identically for the
three magnet setting pairs? If, on the other hand $(\cdot)$ is just
seen as a random variable, why do the functions $A$ not also depend
on a measurement time label, as they are introduced in the theory of
stochastic processes~\cite{BREU02}? \revision{We give below a clear
answer to these questions by means of our counterexample but discuss
first the different views taken in the well established probability
theories of Boole and Kolmogorov as well as quantum mechanical
``probability'' as introduced by the Born rule.}

The probability theory of Boole and its generalization and
perfection by Kolmogorov reduce the actual experiments to logical
abstractions and establish a one to one correspondence between the
experiments and these abstractions. For the case that interests us
we have only two possible experimental outcomes denoted by $\pm 1$
(or equivalently $0, 1$ or $true$ and $false$). ``Probability" is
defined by Boole and Kolmogorov by imposing a measure (a real number
of the interval $[0, 1]$) onto these elements that is consistent
with the experimental factors related to both the single logical
abstractions as well as the whole set of these abstractions. This is
the hallmark of modern probability theory and emphasizes the
relation to set theory.

The one to one correspondence of mathematical abstractions to actual
experiments and a measure on the set of these abstractions are both
necessary to give meaning to the word probability in a set-theoretic
sense. The less familiar reader is encouraged to look at these
definitions in the original work of Boole~\cite{BO1862} or, for the
Kolmogorov framework, in textbooks such as~\cite{FELL68}. For such a
model to make general sense in all experimental situations, we must
assume that (1) a given and well defined logical element
representing an experimental outcome or, in the
language of Kolmogorov, an elementary event will occur with the same
probability measure throughout all experiments and that (2) the
physical characterization of the logical elements of Boole
(elementary events of Kolmogorov) is consistent and complete
throughout the experimental sequence.

This requirement for the description of experiments by mathematical
and logical abstractions that represent a ``truth'' throughout an
experimental sequence, brings us back to the fundamental statement
of Plato's logic: ``$P$ aut non $P$ tertium non datur'' 
%(there are no more than two possibilities: proposition $P$ is true or it is not true) 
and goes to the heart of discussions related to questions such
as ``does the moon shine when I am not looking?". The sentence ``The
moon shines'' is, in general too ill defined to be identified with a
logical variable, say $B$ that assumes a value
$+1$ if the moon shines and $-1$ if it does not. Throughout any
reasonably general experimental sequence that lasts for a certain
duration, the moon may or may not shine at certain different places
and $B$ will therefore assume a variety of values at these different
places. Correspondingly a certain outcome of $B$ can not stand for
the same mathematical abstraction that describes facts at different
locations. If we wish to associate with $B$ a certain truth or
logical expression that is valid everywhere and throughout the
experimental sequence we need to introduce some generalized
coordinates and formulate a more precise statement such as ``the
moon was shining in Monte Carlo at a certain date and time". In
connection with general science experiments we need to note that a
statement about experimental outcomes often may make no sense
whatsoever without the introduction of a coordinate system.

Therefore, we propose the use of the space-time of special
relativity to complete the characterization of Boole's logical
elements and Kolmogorov's elementary events. We assume that only
this completion can lead to $true$-$false$ or other binary statements
that are always and everywhere valid even in very complex one to one
correspondences of mathematical abstractions with
actual experiments.

We can, as a simple example, have a
number of coins and measure the outcome of coin-tosses at certain
given space-time coordinates.
The coins may contain some magnetic material and there may be hidden
magnets with settings ${\bf a}, {\bf b}, {\bf c}$ that co-determine
a probability to measure head or tail for the given coins at the
given space-time coordinates. For given magnet settings and
space-time coordinates of the coins we have then certain outcomes
that form a sample space and certain probabilities for the outcomes
that together with the sample space form a probability
space~\cite{FELL68}. If we do not label the coins by their correct
space-time coordinates then we may have, for example, different
magnet settings applying to the same coin and therefore may have
different probabilities for the outcomes of the coin toss which may
lead to confusion and contradictions.

Quantum theory uses a variation of probability theory by invoking
a wave function $\psi$ that does not have a direct physical
interpretation but does correspond to a certain experimental
procedure of preparation. The settings of the macroscopic
measurement equipment can be chosen at will and the measurements may
be performed involving detection of particles that involve a
space-time
description through the many-body Hamiltonian and wave function
$\psi$. The ``probability" to measure a particle by the given
equipment with given setting is then related by Born's
interpretation to the absolute square (a positive number) of the
wave function that thus assigns a positive number to an event once
the actual type of measurement is chosen. This assignment, however,
can not yet be regarded as a probability measure in the spirit of
Boole or in terms of Kolmogorov's definitions because there is no
assignment made at this point for a sample space, i.e. a space of
all possible outcomes and corresponding elementary events or logical
elements. The Born rule appears thus as a pre-measure that may be
expanded to a full Kolmogorov probability measure only after all
experiments of a sequence are chosen i.e. once all macroscopic
equipment configurations of measurements and all possible outcomes
(data) are fully determined. If we desire to create a Kolmogorov
frame model based on Born's rule, then the actual choice of random
variables may also necessitate the introduction of one or more
stochastic processes in order to include time
coordinates that are otherwise not included in the Kolmogorov
framework. Even this advanced procedure as described e.g.
in~\cite{BREU02} leaves us with the vexing
problem of determination which mathematical abstractions (elementary
events of Kolmogorov or logical elements of Boole) correspond to the
different actual experiments.

For example, assume that one measures correlated pairs of spin $1/2$
particles with magnet settings ${\bf a}, {\bf b}$ and characterizes
the dichotomic outcomes for the ${\bf a}, {\bf b}$ settings by the
variables $A_{\bf a}, A_{\bf b}$. Further assume that in another set
of measurements we measure with magnet settings ${\bf a}, {\bf c}$.
Can we then denote the corresponding variables for the outcomes by
$A_{\bf a}, A_{\bf c}$? Recall that, in this second case, we measure
the ``$A_{\bf a}$'' outcomes corresponding to the $\bf c$ setting
(in the other wing of the experiment) at different space-time
coordinates and with different correlated pairs as compared to the
first case ``$A_{\bf a}$'' outcomes that correspond to the original
$\bf b$ setting. Is it then permitted to use the same dichotomic
variable or logical element as used for the $\bf b$ setting?

Because a sample space and single outcomes are not included into
considerations of quantum theory, this theory does not answer the
above question. The Born rule per se does therefore not provide
probabilities in the sense of Boole or Kolmogorov but can only lead
to a probability once a one to one assignment of mathematical
elements and experimental outcomes is made and a measure
for the whole space of possible outcomes, the whole
sample space, is introduced. This can not be accomplished by
normalizing a given wave function because that normalization refers
only to a single preparation and measurement of a much more
elaborate sequence of experiments. However, it is clear that for
measurements with a given macroscopic setting and a fixed method of
preparation, sample spaces can always be created and that such a
sample space of measurement outcomes together with the probabilities
from Born's rule forms then also a probability space \`{a} la
Kolmogorov for a given setting as outlined in texts such
as~\cite{BREU02}. Nevertheless, for different and particularly for
incompatible experiments and for a given characterization of
functions or random variables e.g. by magnet settings only, such a
probability space may not exist. 

As we will see in our
%counterexample this non-existence depends crucially on the
%space-time characterization (or non-characterization) of the
%experimental outcomes and their one to one correspondence to the
%mathematical idealizations be they elements of Boolean logic or
%elementary events in the framework of Kolmogorov.
counterexample this non-existence depends crucially on the
one-to-one correspondence of the experimental outcomes to their 
mathematical idealizations be they elements of Boolean logic or
elementary events in the framework of Kolmogorov.

Many mathematical papers on probability theory
simply start with the phrase ``given a Kolmogorov probability
space...''. It is, however, well known and has been particularly
well pointed out by Vorob'ev~\cite{VORO62} that there are cases in
which a Kolmogorov probability space does not exist. In particular,
there exist numerous classical experiments that subject to certain
characterizations by simple settings, can not be described on one
probability space in a logically consistent way. Take, for example,
certain physical experiments that can be described by Stochastic
Processes. Examples are Brownian motion or stock market and exchange
rate fluctuations. It is plausible that such different processes may
not be describable by a single stochastic process but are described
rather by different ones. It is less known but has been shown in
great detail that even very slight changes in experiments may
require the use of different stochastic processes for their
description and that this is true also for EPR-type
experiments.%~\cite{HESS08}.

It is the purpose of this paper to show that Born's rule defines a
pre-probability measure that only then can be turned into a
Kolmogorov (or Boole) probability if a logically consistent one to
one correspondence between experimental outcomes and mathematical
abstractions is or can be made. We also show that such one to one
correspondence can always be made for the known EPR experiments by
completing the characterization of the mathematical symbols
describing the functions $A$ of Bell by use of space-time indices
that relativity theory provides us with. Indices related to
influences at a distance would also accomplish the same goal of
obtaining a consistent probability measure \`{a} la Kolmogorov from
Born's rule but do not appear to be necessary.

\section{Games with symptoms and patients: From Boole to Bell}

As mentioned, the early definitions of probability by Boole were
related to a one to one correspondence that Boole established
between actual experiments and idealizations of them through
elements of logic with two possible outcomes. His view gave the
concept of probability precision in its relation to sets of
experiments and this precision is expressed by Boole's discussion of
probabilities as related to possible experience.
These discussions can be best explained by an example
that also shows the role of space-time coordinates in
the characterization of variables related to probability theory. We
discuss first this example that has its origins in the works of
Boole and also Vorob'ev and relates to the work of Bell inasmuch as
it can be used as a counterexample to Bell's conclusions related to
non-locality. Then we return to the more general discussions of
probability in quantum theory.

Consider a certain disease that strikes persons in different ways
depending on circumstances such as place of birth and place of
residence etc.. Assume that we deal with 
\revision{one set} 
of patients that are born in Africa (subscript $\bf a$), in Asia
(subscript $\bf b$) and in Europe (subscript $\bf c$). 
Assume further that
doctors are assembling information about the disease altogether in
the three cities Lille, Lyon and Paris, all in France. The doctors
are careful and perform the investigations on randomly chosen but
identical dates. The patients are denoted by the symbol $A_{\bf
o}^l(n)$ where ${\bf o} = {\bf a}, {\bf b}, {\bf c}$ depending on
the birthplace of the patient, $l = 1, 2, 3$ depending on where the
doctor gathered information $1$ designating Lille, $2$ Lyon and $3$
Paris respectively, and $n = 1, 2, 3,\ldots,N$ denotes just a given
random day of the examination. The doctors assign a value $A = \pm
1$ to each patient; $A = +1$ if the patients show a certain symptom
and $A = -1$ if they do not.

The first variation of this investigation of the disease is
performed as follows. 
The doctor in Lille examines 
\revision{all} 
patients of type $\bf a$,
the doctor in Lyon 
\revision{all} 
patients of type $\bf b$
and the doctor in Paris 
\revision{all} 
patients of type $\bf c$.
On any given day of examination (of
precisely one patient for each doctor and day) they write down their
diagnosis and then, after many exams, concatenate the results and
form the following sum of pair-products of exam outcomes at a given
date described by $n$:
\begin{equation}
\Gamma(n) = A_{\bf a}^1(n)A_{\bf b}^2(n) + A_{\bf a}^1(n)A_{\bf
c}^3(n) + A_{\bf b}^2(n)A_{\bf c}^3(n)
.
\label{hla23n1}
\end{equation}
Boole noted now that
\begin{equation}
\Gamma(n) \geq -1
,
\label{hla23n2}
\end{equation}
which can be found by inserting all possible values for the patient
outcomes summed in Eq.~(\ref{hla23n1}). For the average (denoted by
$\langle . \rangle$) over all examinations we have then also:
\begin{equation}
\Gamma= \langle\Gamma(n)\rangle=\frac{1}{N}\sum_{n=1}^N \Gamma(n)
\geq -1
.
\label{hla23n3}
\end{equation}
This equation gives conditions for the product averages and
therefore for the frequencies of the concurrence of certain values
of $A_{\bf a}^1(n), A_{\bf b}^2(n)$ etc. e.g. for $A_{\bf a}^1(n)
=+1, A_{\bf b}^2(n) = -1$. These latter frequencies must therefore
obey these conditions. Thus we obtain rules or non-trivial
inequalities for the frequencies of concurrence of the patients
symptoms. Boole calls these rules ``conditions
of possible experience". In case of a violation, Boole states that
then the ``evidence is contradictory''.

In the opinion of the authors, the term ``possible experience'' is
somewhat of a misnomer. The experimental outcomes have been
determined from an experimental procedure in a scientific way and
are therefore possible. What may not be possible is the one to one
correspondence of Boole's logical elements or variables to the
experimental outcomes that the scientist or statistician has chosen.
In order to judge precisely where the contradictions arise from, we
need to advance 100 years to the work of Vorob'ev on the one side
and go back to the meaning of Plato's logic and his rule ``aut $P$
aut non $P$ tertium non datur" on the other.

Before doing so, however, we note the following. In this example, we
may indeed regard the various $A_{\bf o}^l(n) = \pm1$ with given
indices as the elements of Boole's logic to which the actual
experiments can be mapped. As shown by Boole, this is a sufficient
condition for the inequality of Eq.~(\ref{hla23n3}) to be valid. We
may in this case also omit all the indices except for those
designating the birth place and still will obtain a valid equation
that can never be violated:
\begin{equation}
\langle A_{\bf a}A_{\bf b}\rangle + \langle A_{\bf a}A_{\bf c}\rangle + \langle A_{\bf
b}A_{\bf c}\rangle \geq -1
.
\label{hla23n3b}
\end{equation}
The reason is simply that three arbitrary dichotomic variables i.e.
variables that assume only two values ($\pm 1$ in our case) must
always fulfill Eq.~(\ref{hla23n3b}) no matter what their logical
connection to experiments is because we deduce the three products of
Eq.~(\ref{hla23n3b}) from sequences of each three measurement
outcomes. Note that Eq.~(\ref{hla23n3b}) contains six factors with
each birthplace appearing twice and representing then the
identical result. Below we will discuss a slightly
modified experiment that is much more general and contains six
measurement results for the six factors. Before discussing this more
general experiment that resembles more clearly EPR experiments we
turn now to the findings of Vorob'ev regarding this type of
inequalities and Boole's conditions of possible experience.

Obviously the inequality of Eq.~(\ref{hla23n2}) is non-trivial
because based on the fact that the value of all
products must be $\pm 1$ one could only conclude that
\begin{equation}
\Gamma(n) \geq -3
.
\label{hla23n4}
\end{equation}
The nontrivial result has the following reason. Boole included into
Eq.~(\ref{hla23n1}) a cyclicity: the outcomes of the first two
products determine the outcomes in the third product. Because all
outcomes can only be $\pm 1$ the cyclicity gives rise to
Eq.~(\ref{hla23n2}). Vorob'ev showed
precisely 100 years after Boole's original work in a very general
way that it is always a combinatorial-topological cyclicity that
gives rise to non-trivial inequalities for the mathematical
abstractions of experimental outcomes. Boole pointed to the fact
that Eq.~(\ref{hla23n2}) can not be violated.
However, in order to come to that conclusion, the $A_{\bf o}^l(n)$
need, in the first place, to be in a one to one
correspondence to Boole's elements of logic that follow the law
``aut $A = +1$ aut $A = -1$ tertium non datur". As discussed in the
introduction, eternally valid statements about physical experience
such as ``aut $A = +1$ aut $A = -1$ tertium non datur" can usually
not be made when describing the physical world without the use of
some coordinates. In the example above these coordinates where the
places of birth, the places of examination and the numbering of the
exams that were randomly taken. All these coordinates when added
need to still allow for a cyclicity in order to make Boole's
inequality non-trivial. Therefore, if we have a violation of a
non-trivial Boole inequality, then we must conclude that we have not
achieved a one to one correspondence of our variables to the
elementary eternally true logical variables of Boole and that we
need further ``coordinates'' that will then remove the cyclicity. In
order to illustrate all this by a simple example, we consider the
following second different statistical investigation of the same
disease.

We now let only two doctors, one in Lille and one in Lyon perform
the examinations. The doctor in Lille examines randomly 
\revision{all} patients of
types $\bf a$ and $\bf b$ and the one in Lyon \revision{all} of type $\bf b$ and
$\bf c$ each one patient at a randomly chosen date. 
\revision{Note that in this way, all patients of type $\bf b$
receive two examinations.}
The doctors are
convinced that neither the date of examination nor the location
(Lille or Lyon) has any influence and therefore denote the patients
only by their place of birth. After a lengthy period of examination
they find:
\begin{equation}
\Gamma = \langle A_{\bf a}A_{\bf b}\rangle + \langle A_{\bf a}A_{\bf c}\rangle + \langle A_{\bf
b}A_{\bf c}\rangle = -3 .
\label{hla23n5}
\end{equation}
They further notice that the single outcomes of $A_{\bf a}, A_{\bf
b}$ and $A_{\bf c}$ are randomly equal to $\pm 1$. This latter fact
completely baffles them. How can the single outcomes be entirely
random while the products are not random at all and how can a Boole
inequality be violated hinting that we are not dealing with a
possible experience? After lengthy discussions they conclude that
there must be some influence at a distance going on and the outcomes
depend on the exams in both Lille and Lyon such that a single
outcome manifests itself randomly in one city and that the outcome
in the other city is then always of opposite sign. Naturally that
way they have removed the Vorob'ev cyclicity and we have only the
trivial inequality Eq.~(\ref{hla23n4}) to obey.

However, there are
also other ways that remove the cyclicity, ways that do not need to
take recourse to influences at a distance. For example we can have a
time dependence and a city dependence of the illness as follows. On
even dates we have $A_{\bf a} = +1$ and $A_{\bf c} = -1$ in both
cities while $A_{\bf b} = +1$ in Lille and $A_{\bf b} = -1$ in
Lyon. On odd days all signs are reversed. Obviously for
measurements on random dates we have then the outcome that $A_{\bf
a}, A_{\bf b}$ and $A_{\bf c}$ are randomly equal to $\pm 1$ while
at the same time $\Gamma(n) = -3$ and therefore $\Gamma = -3$. We
need no deviation from conventional thinking to arrive at this
result because now, in order to deal with Boole's elements of logic,
we need to add the coordinates of the cities to obtain
$
%\begin{equation}
\Gamma = \langle A_{\bf a}^1 A_{\bf b}^2\rangle + \langle A_{\bf a}^1 A_{\bf c}^2\rangle +
\langle A_{\bf b}^1 A_{\bf c}^2\rangle \geq -3
%,
%\label{hla23n6}
%\end{equation}
$ and the inequality is of the trivial kind because the cyclicity is
removed. The date index does not matter for the products since both
signs are reversed leaving the products unchanged. However, in
actual fact, also this index might have to be included and could be
a reason to remove the cyclicity e.g.
%
%\begin{eqnarray}
%\Gamma = \langle A_{\bf a}^1(d_1) A_{\bf b}^2(d_1)\rangle &+& \langle A_{\bf a}^1(d_2)
%A_{\bf c}^2(d_2)\rangle
%\nonumber \\
%&+& \langle A_{\bf b}^1(d_3) A_{\bf c}^2(d_3)\rangle \geq -3
%,
%\label{hla23n6a}
%\end{eqnarray}
$
\Gamma = \langle A_{\bf a}^1(d_1) A_{\bf b}^2(d_1)\rangle + \langle A_{\bf a}^1(d_2)
A_{\bf c}^2(d_2)\rangle
+ \langle A_{\bf b}^1(d_3) A_{\bf c}^2(d_3)\rangle \geq -3
$,
where we now have included the fact that the exams of pairs are
performed at different dates $d_1, d_2, d_3$.

We note that in connection with EPR experiments and questions
relating to interpretations of quantum theory,
Eqs.~(\ref{hla27n1}) and (\ref{hla23n2}) are called Leggett-Garg
inequalities and are of the Bell-type. It is often claimed that a
violation of such inequalities implies that either realism or
Einstein locality should be abandoned. As we saw in our
counterexample which is both Einstein local and realistic in the
common sense of the word, it is the one to one correspondence of the
variables to the logical elements of Boole that matters when we
determine a possible experience, but not necessarily the choice
between realism and Einstein locality. Phrased differently, the
question ``does the moon shine when we are not looking" is simply
too imprecise. Had we given a space-time coordinate for the event
that the moon shines we would have expressed an eternal truth of a
measurement.

Realism plays a role in the arguments of Bell and followers because
they introduce a variable $\lambda$ representing an element of
reality and then write
\begin{equation}
\Gamma = \langle A_{\bf a}(\lambda) A_{\bf b}(\lambda)\rangle + \langle A_{\bf
a}(\lambda) A_{\bf c}(\lambda)\rangle + \langle A_{\bf b}(\lambda) A_{\bf
c}(\lambda)\rangle \geq -1
.
\label{hla23n7}
\end{equation}
Because no $\lambda$ exists that would lead to a violation except a
$\lambda$ that depends on the index pairs ($\bf a$, $\bf b$), ($\bf
a$, $\bf c$) and ($\bf b$, $\bf c$) the simplistic conclusion is
that either elements of reality do not exist or they are non-local.
The mistake here is that Bell and followers insist from the start
that the same element of reality occurs for the three different
experiments with three different setting pairs. This assumption
implies the existence of the combinatorial-topological cyclicity
that in turn implies the validity of a non-trivial inequality but
has no physical basis. Why should the elements of reality not all be
different? Why should they, for example not include the time of
measurement? There is furthermore no reason why there should be no
parameter of the equipment involved. Thus the equipment could
involve time and setting dependent parameters such as $\lambda_{\bf
a}(t), \lambda_{\bf b}(t), \lambda_{\bf c}(t)$ and the functions $A$
might depend on these parameters as well. \revision{We refer the
reader to the
references~\cite{FINE82,HESS01,RAED06c,ZHAO08,KHRE09} and note that
parameters related to the devices of measurement have been discussed
already by Wigner~\cite{WIGN70} but not in connection to
the one-to-one correspondence with Boole's logical elements.
The possible dependence of these parameters on
measurement time or Einstein's space-time prevents the derivation of
the Clauser-Horne-Shimony-Holt inequality because outcome
independence may be violated, as can be seen directly by using our
example of different outcomes for even and odd times in their
equations.}

\section{Bell revisited from the view of quantum theory}

Consider three spin 1/2 particles that are measured by macroscopic
equipment involving three Stern-Gerlach magnets. The wave function of
the three particles is not nearer specified and denoted by $\psi_3$.
If we denote the measurement outcomes at measurement time $n$
for the three particles with
the three respective magnet settings by $A_{\bf a}^n(\psi_3), A_{\bf
b}^n(\psi_3), A_{\bf c}^n(\psi_3)$, then it is easy to show by the
laws of quantum theory that the Boole (Bell) inequality~\cite{RAED09a}:
\begin{eqnarray}
\langle A_{\bf a}^n(\psi_3)A_{\bf b}^n(\psi_3)\rangle &+& \langle A_{\bf
a}^n(\psi_3)A_{\bf c}^n(\psi_3)\rangle
\nonumber \\
&+& \langle A_{\bf b}^n(\psi_3)A_{\bf
c}^n(\psi_3)\rangle  \geq -1 \label{hlm2n1}
,
\end{eqnarray}
is fulfilled and we can conclude that we have dealt with the logical
elements of Boole and well defined probabilities.

If we consider instead six measurements of pairs of particles that are
described by the singlet state $\psi_S$ then we need three different
measurement station pairs or one pair of measurement-stations at
three different measurement times. For simplicity consider three
different measurement-station pairs that we label with indices $n,
m, l$. Correspondingly we also introduce for the measurement
outcomes the symbols $A_{\bf a}^n(\psi_S), A_{\bf b}^n(\psi_S);
A_{\bf a}^m(\psi_S), A_{\bf c}^m(\psi_S); A_{\bf b}^l(\psi_S),
A_{\bf c}^l(\psi_S)$. Then quantum theory tells us that for
certain magnet settings we may have:
\begin{eqnarray}
\langle A_{\bf a}^n(\psi_S)A_{\bf b}^n(\psi_S)\rangle
&+& \langle A_{\bf a}^m(\psi_S)A_{\bf c}^m(\psi_S)\rangle
\nonumber \\
&+& \langle A_{\bf b}^l(\psi_S)A_{\bf c}^l(\psi_S)\rangle <  -1 \label{hlm2n2}
,
\end{eqnarray}
and we have a violation of an inequality that resembles the Bell-
type. In this case, however, this does not surprise us because as
long as we have no cyclicity in the expressions of
Eq.~(\ref{hlm2n2}), we obtain only a trivial Boole inequality and as
far as Boole's or Kolmogorov's probability are concerned the right
hand side of Eq.~(\ref{hlm2n2}) might as well be $-3$. Note that the
attachment of space-time indices to the variables that provide a
characterization of the experiments in addition to observations such
as the magnet settings always permit a removal of any cyclicity.
Quantum theory does not have any concerns about the indices $n,
m, l$ because quantum theory is careful not to assign any meaning
to the single outcomes and therefore does not rely on or need a
sample space or probability space.

A probability as in the frameworks of Boole or Kolmogorov is thus
not defined in quantum theory because quantum theory does not
define any relations of its framework to single logical elements or
elementary events and therefore also can not provide a measure to
general sets or subsets of such elements or events. What is defined
in quantum theory are long term averages and these may be related in
a variety of ways to the actual logical elements of a theory. The
probability amplitude just carries with it all the possibilities
that may actually be realized in a set of data, that is all the
possibilities that may be realized as a sample space. For an actual
sample space to be realized other choices must be made that, in
principle, have nothing to do with the quantum particles that are
measured but only with the macroscopic equipment that is brought
into a certain setting for the purpose of measurement. These other
choices may again involve sample spaces and probability spaces that
together with the measurement outcomes related to quantum particles
may form complex stochastic processes.

Quantum theory predicts the long term averages of these
stochastic processes but does not attempt to unify these processes
into one common stochastic process. The Born rule thus attaches
positive values to measurement outcomes that are related to certain
measurements and preparations and defines in this way what one could
call a pre-measure. For all well defined macroscopic equipment
arrangements this pre-measure can be turned into a probability
measure with different experimental sequences corresponding, in
principle, to different probability measures. Whether or not these
different measures and sample spaces can be unified is a matter of
characterization. If no unification is possible, as would be
indicated by a violation of a Boole (Bell) inequality, then one
needs further detail in the characterization of variables in order
to remove the cyclicity. That may be achieved both in an Einstein
local way or in a non-local fashion. As we saw above, EPR
experiments always permit extended characterization by Einstein's
space-time and corresponding avoidance of cyclicity. Nonlocal
characterizations that avoid cyclicity are also always possible but
not necessary. The only alternative to the above is to abandon
realism (whatever we mean by this word) altogether. The examples
(counterexamples) with the patient-investigations and the relation
of these examples to EPR experiments prove, at least in the opinion
of these authors, that neither realism nor Einstein locality need be
abandoned because of a violation of Bell's inequalities.

\bibliographystyle{eplbib}
\bibliography{../epr}

\end{document}